\documentclass[aps,prb,groupedaddress,showpacs,twocolumn,superscriptaddress,10pt]{revtex4-2}

\usepackage{graphicx}
\usepackage{mathtools}
\usepackage{amsmath}
\usepackage{amssymb}
\usepackage{bbm}
\usepackage{bm}
\usepackage{cancel}
\usepackage{xspace}
\usepackage{braket}
\usepackage{xcolor}   
\usepackage{calrsfs}
\usepackage{comment}
\usepackage{placeins}
\usepackage[colorlinks=true]{hyperref} 

\usepackage{units}
 
\DeclareMathAlphabet{\pazocal}{OMS}{zplm}{m}{n} 

\newcommand{\ee}{\mathrm{e}}  
\DeclareMathOperator*{\ii}{i} 
\newcommand*\dd{\mathop{}\!\mathrm{d}}

\renewcommand{\vec}[1]{\bm{#1}} 
\newcommand{\mat}[1]{\bm{#1}} 
\newcommand{\kel}[1]{\underline{#1}} 

\DeclareMathOperator*{\rre}{Re}
\DeclareMathOperator*{\iim}{Im}
\DeclareMathOperator*{\Tr}{Tr}

\newcommand{\aamp}{\mathcal{A}}

\definecolor{orange}{RGB}{252,77,6}
\definecolor{brown}{RGB}{200,127,50}
\definecolor{blue}{RGB}{0,0,255}

\definecolor{ao(english)}{rgb}{0.0, 0.5, 0.0}

\definecolor{hblue}{RGB}{0,0,255}

\begin{document}

\title{Impact ionization processes in a photodriven Mott insulator: influence of phononic dissipation} 

\author{Paolo Gazzaneo}
\email[]{paolo.gazzaneo@tugraz.at}
\affiliation{Institute of Theoretical and Computational Physics, Graz University of Technology, 8010 Graz, Austria}
\author{Tommaso Maria Mazzocchi}
\affiliation{Institute of Theoretical and Computational Physics, Graz University of Technology, 8010 Graz, Austria}
\author{Jan Lotze}
\affiliation{Institute of Theoretical and Computational Physics, Graz University of Technology, 8010 Graz, Austria}
\author{Enrico Arrigoni}
\email[]{arrigoni@tugraz.at}
\affiliation{Institute of Theoretical and Computational Physics, Graz University of Technology, 8010 Graz, Austria}

\date{\today} 
 
\begin{abstract}     
 
We study a model for photovoltaic energy collection consisting of a Mott insulating layer in presence of acoustic phonons, coupled to two wide-band fermion leads at different chemical potentials and driven
into a nonequilibrium steady state by a periodic electric field. We treat electron correlations with nonequilibrium dynamical mean-field theory (DMFT) using the so-called auxiliary master equation approach as impurity solver and include dissipation by acoustic phonons via the Migdal approximation. For a small hybridization to the leads, we obtain a peak in the photocurrent as a function of the driving frequency which can be associated with impact ionization processes.
For larger hybridizations the shallow peak suggests a suppression of impact ionization with respect to direct photovoltaic excitations. Acoustic phonons slightly enhance the photocurrent for small driving frequencies and suppress it at frequencies around the main peak at all considered hybridization strengths.

\end{abstract} 

       
\maketitle

\section{Introduction} 
\label{sec:intro}

The idea of designing photovoltaic devices exploiting the Mott gap to convert electromagnetic radiation into energy has become popular in recent years~\cite{mano.10,li.ch.13,gu.gu.13,co.ma.14,wa.li.15}. In particular, it has been suggested that in strongly correlated materials, highly excited charge carriers could use their extra energy to excite additional carriers across the Mott gap via impact ionization (II)~\cite{mano.10,co.ma.14}, thus potentially improving their efficiency beyond the Shockley-Queisser limit~\cite{sh.qu.61}. Although II is also present in conventional semiconductor devices, the time scales for electron-electron scattering are typically much longer than in correlated materials, so that highly excited electrons will mostly dissipate their energy to phonons. 

Experimentally, evidence for fast carrier multiplication processes has been detected via pump-probe experiments in VO$_2$~\cite{ho.bi.16}. Oxide heterostructures based on LaVO$_3$/SrTiO$_3$ have been identified as promising candidates, due to the ideal band gap and the strong polar field, able to separate the excited charge carriers~\cite{as.bl.13}. There are however some drawbacks, such as the low mobility of the carriers~\cite{wa.li.15,je.re.18}, which still put into question these materials’ applicability as efficient solar cells. Thus while a large-scale application of Mott-based solar cells may be difficult to achieve, employing them as photodetectors may be more promising on the long run, due to their high photoresponsivity. From the scientific point of view and for future applications it is thus worthwile to further investigate the properties of Mott-based photovoltaic materials and the mechanisms behind their inner workings, to explore whether alternative unexpected paths can be opened. In fact, a lot of theoretical work has been done to understand the II process (see, e.g.~\cite{co.ma.14,ec.we.11,ec.we.13,we.he.14,pe.be.19,so.do.18,ka.wo.20,mano.19,ma.ev.22}).

As a scattering process, II competes with electron-phonon (e-ph) scattering which is the dominant relaxation mechanism in conventional semiconductors. This may not be the case in Mott insulators, at least in certain cases~\cite{co.ma.14}. Therefore, it is essential to understand the influence of e-ph scattering on the photocurrent and II in Mott photovoltaic devices, which is the issue we address in this paper.

\begin{figure}[b] 
\includegraphics[width=0.8\linewidth]{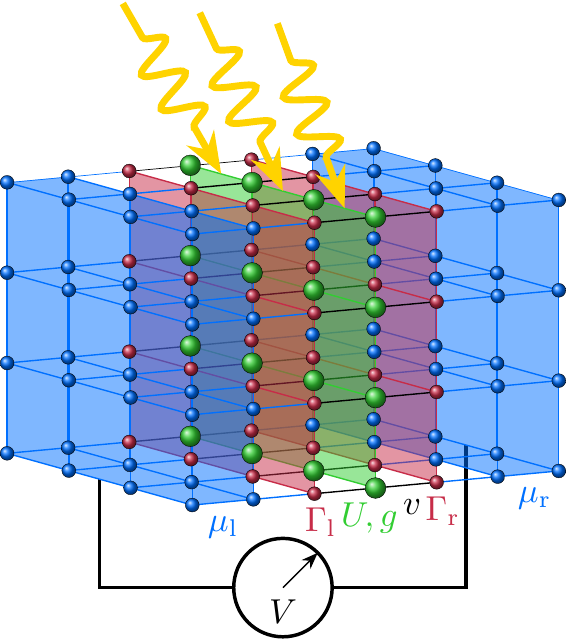}
\caption{(Color online) Schematic representation of the setup. A central, correlated layer with local Hubbard interaction $U$ and e-ph coupling $g$ (green) is sandwiched between two non-interacting layers (red) via the hybridization $v$ (black). The latter are in turn coupled to wide-band fermion reservoirs (blue) with different chemical potentials $\mu_{\text{l}/\text{r}}$, which introduce broadenings $\Gamma_{\text{l}/\text{r}}$.}
\label{fig:setup}
\end{figure}

For this goal we consider the simplified model shown in Fig.~\ref{fig:setup}, consisting of a Hubbard layer located between two leads. The leads have themselves a wide band, as expected for good conductors, which is effectively narrowed due to the weak coupling to intermediate layers which we take as simplified models for the contacts. Electrons on the Hubbard layer interact locally with acoustic phonons. An external periodic electric field induces a Floquet steady state with a current flowing from the lead with the lower to the lead with the higher chemical potential, thus transferring its energy. We address this periodic problem via nonequilibrium Floquet dynamical mean-field theory (F-DMFT).            
        
Our goal is to investigate the occurrence of II in this photovoltaic setup and how it is affected by the presence of phonons. Specifically, we study the behavior of the photovoltaic current, the spectral properties as well as the double occupation as a function of the hybridization strength and the properties of the phonons. Our analysis can only provide qualitative results, while quantitative comparisons with experimental systems require a more realistic setup and are beyond the scope of this work.
 
Our calculations show that a small hybridization to the leads is more favorable for II than a larger one. II therefore plays a crucial role in compensating the smaller charge injection, giving the same order of magnitude in the photocurrent as for the larger hybridization.
In all cases, the interaction with phonons slightly enhances the photocurrent for small driving frequencies, while suppressing it in vicinity of the main peak.

The structure of the paper is as follows: in Sec.~\ref{sec:Model} we describe the model Hamiltonian. Methods, formalism and observables are presented in Sec.~\ref{sec:Method_formalism}. We discuss our results in Sec.~\ref{sec:results} and present our conclusion in Sec.~\ref{sec:conclusions}. 

\section{Model}
\label{sec:Model}

Our setup introduced in Sec.~\ref{sec:intro} and shown in Fig.~\ref{fig:setup} is described by the Hamiltonian
\begin{equation}
\label{eq:Hamiltonian}
\begin{split} 
\hat{H}(t)&=\varepsilon_{\text{c}} \sum_{i\sigma}\hat{n}_{i\sigma} -\sum_{\sigma}\sum_{(i,j)} t_{ij}(t) \hat{c}^{\dagger}_{i\sigma} \hat{c}_{j\sigma}
\\ & + U \sum_{i} \hat{n}_{i\uparrow} \hat{n}_{i\downarrow}+\hat{H}_{\text{e-ph}} + \hat{H}_{\text{ph}}+\hat{H}_{\text{lead}}.
\end{split}
\end{equation}
The operator $\hat{c}^{\dagger}_{i\sigma}$ ($\hat{c}_{i\sigma}$) creates (annihilates) an electron with spin $\sigma= \lbrace \uparrow,\downarrow \rbrace$ on the $i$-th lattice site and $\hat{n}_{i\sigma}\equiv \hat{c}^{\dagger}_{i\sigma} \hat{c}_{i\sigma}$ is the particle number operator. We denote with $(i,j)$ the sum over nearest-neighbor sites and with $\varepsilon_{\text{c}} \equiv -U/2$ the on-site energy. The driving field consists of a time-periodic, homogeneous and monochromatic electric field with frequency $\Omega$. In the temporal gauge it enters via the Peierls substitution in the time-dependent hopping in Eq.~\eqref{eq:Hamiltonian}~\cite{peie.33}
\begin{equation}\label{eq:peierls} 
t_{ij}(t) = t_{\text{c}} \ e^{-\ii \frac{q}{\hbar} \left( \vec{r}_j - \vec{r}_i \right) \cdot \vec{A}(t)}, 
\end{equation}
where $t_{\text{c}}$ is the intra-layer hopping, $\vec{A}$(t) the time-dependent vector potential, $\hbar$ Planck's constant and $q$ the charge of the electrons. Following Refs.~\cite{ts.ok.08,mu.we.18}, we choose for simplicity $\vec{A}(t)=\vec{e}_{0} A(t)$, which points along the lattice body diagonal $\vec{e}_{0}=(1,1,\dots,1)$. Therein $A(t)=\frac{\hbar}{qa}\aamp\sin(\Omega t)$ with $\aamp=-\frac{qE_0a}{\hbar\Omega}$, $E_0$ is the electric field amplitude, and $a$ the lattice spacing~\cite{ts.ok.08}. 
In this temporal gauge the electric field is $\vec{E}= -\partial_{t}\vec{A}(t) = \vec{e}_{0}E_0 \cos(\Omega t)$.

As in Ref.~\cite{ma.ga.22}, phonons are included in the form of acoustic  branches attached to each lattice site and coupled to the electrons via the Hamiltonian 
\begin{equation}\label{eq:e-ph_ham} 
\hat{H}_{\text{e-ph}} = g \sum_{i\sigma} \hat{n}_{i\sigma} \hat{x}_{i},
\end{equation}
with $g$ as e-ph coupling and $\hat{x}_{i} \equiv \frac{1}{\sqrt{2}} \left( \hat{b}^{\dagger}_{i} + \hat{b}_i \right)$, where $\hat{b}^{\dagger}_{i}$ ($\hat{b}_{i}$) creates (annihilates) a phonon of the acoustic branch at site $i$, with dispersion relation described by $\hat{H}_{\text{ph}}$, discussed in Sec~\ref{sec:Dyson_equation}. In contrast to Holstein phonons, they have a finite bandwidth, which is more effective in energy dissipation. Details on the leads are provided in Sec.~\ref{sec:Dyson_equation}.    
 
For mathematical simplicity~\cite{ge.ko.92,ao.ts.14}, we consider the correlated central region as a $d$-dimensional layer and consider the limit of infinite dimensions $d \rightarrow \infty$ by rescaling the hopping as $t_{\text{c}}=t^{\ast}/(2\sqrt{d})$. Taking $a=1$, sums over the crystal momentum $\vec{k}$ of a generic quantity $\chi$ read $\sum_{\vec{k}} \chi(\omega,\vec{k}) \rightarrow \int \dd\epsilon \int   \dd\overline{\epsilon} \ \rho(\epsilon,\overline{\epsilon}) \chi(\omega;\epsilon,\overline{\epsilon})$ with $\rho(\epsilon,\overline{\epsilon}) = (1/\pi t^{\ast 2}) \exp[-( \epsilon^{2} + \overline{\epsilon}^{2})/t^{\ast 2}]$ as the joint density of states~\cite{ts.ok.08} and with
\begin{align}\label{eq:d-dim_crystal_dep}
\begin{split}
\epsilon & = -2t_{\text{c}} \sum_{i=1}^{d} \cos(k_i a), \\  
\overline{\epsilon}& = -2t_{\text{c}}\sum_{i=1}^{d} \sin(k_i a). \\
\end{split}  
\end{align}
  
Within the remainder of this paper we choose our units so that  $\hbar = k_{\text{B}} = a = 1 = -q$, $t^{\ast}=2\sqrt{2}$~\footnote{With such choice for $t^{\ast}$ we reproduce the DOS of a 2D layer with $t_{\text{c}}=1$  in order to compare with Ref.~\cite{so.do.18}.}, and $t_{\text{resc}} \equiv \frac{t^{\ast}}{2\sqrt{2}}=1$ as unit of energy.

\section{Method and formalism}    
\label{sec:Method_formalism}  

\subsection{Floquet Green's Function method}   
\label{sec:GFs_Dyson_Floquet}

To correctly describe the nonequilibrium periodic steady state, we use the Floquet generalization of the nonequilibrium Green's function (NEGF) approach~\cite{ts.ok.08,sc.mo.02u,jo.fr.08}. Every function $G(t,t^{\prime})$ satisfying the periodicity relation $G(t,t^{\prime})=G(t+\tau,t^{\prime}+\tau)$, with $\tau=2\pi/\Omega$ as the period related to the external driving frequency $\Omega$, may be represented as~\cite{ts.ok.08,sc.mo.02u,jo.fr.08}
\begin{equation}
\label{eq:FloquetGF} 
\kel{G}_{mn}(\omega) =\int \dd t_{\text{rel}} \int_{-\tau/2}^{\tau/2} \frac{\dd t_{\text{av}}}{\tau} \ee^{\ii[\left(\omega+m\Omega\right) t -\left( \omega+n\Omega\right)t^{\prime}]} \kel{G}(t,t'),
\end{equation}
known as the \emph{Keldysh-Floquet} GF. The integration variables $t_{\text{rel}} = t-t^{\prime}$ and $t_{\text{av}} = (t+t^{\prime})/2$ are the relative and average times. Within the rest of this work, we denote a Floquet-represented matrix either as $X_{mn}$ with explicit indices, or use a boldface letter $\mat{X}$. The underline indicates the overall \emph{Keldysh} structure
\begin{equation}\label{eq:Keld-structure}
\kel{\mat{G}} \equiv
\begin{pmatrix}
\mat{G}^{\text{R}} & \mat{G}^{\text{K}}\\
\mat{0}            & \mat{G}^{\text{A}} \\
\end{pmatrix},
\end{equation}
which contains the \emph{retarded}, \emph{advanced} and \emph{Keldysh} components $\mat{G}^{\text{R,A,K}}$, where $\mat{G}^{\text{A}}=(\mat{G}^{\text{R}})^{\dagger}$. The \emph{Keldysh} component is defined as $\mat{G}^{\text{K}} \equiv \mat{G}^{>} + \mat{G}^{<}$, with $\mat{G}^{\lessgtr}$ as \emph{lesser} and \emph{greater} components~\cite{schw.61,keld.65,ra.sm.86,ha.ja}.
  
\subsection{Dyson equation} 
\label{sec:Dyson_equation}

The lattice Floquet GF of our setup obeys the Dyson equation 
\begin{equation}\label{eq:FullDysonEq}
\kel{\mat{G}}^{-1}(\omega_n;\epsilon,\overline{\epsilon}) = \kel{\mat{G}}^{-1}_{0}(\omega_n;\epsilon,\overline{\epsilon}) - \kel{\mat{\Sigma}}(\omega_n;\epsilon,\overline{\epsilon}) - \kel{\mat{\Sigma}}_{\text{e-ph}}(\omega_n;\epsilon,\overline{\epsilon}),
\end{equation}
where in our approximation the electron self-energy (SE) $ \kel{\mat{\Sigma}}$ and e-ph SE $\kel{\mat{\Sigma}}_{\text{e-ph}}$ contribute independently to the total SE~\footnote{In fact, there is an indirect feedback via the DMFT self-consistency.}. Therein, the lattice GF of the non-interacting part of the Hamiltonian in Eq.~\eqref{eq:Hamiltonian} is 
\begin{equation}\label{eq:non-int_InvGF} 
\kel{G}^{-1}_{0,mn}(\omega_n;\epsilon,\overline{\epsilon}) =  \kel{g}^{-1}_{0,mn}(\omega_n;\epsilon,\overline{\epsilon}) - \sum_{\rho\in\{\text{l},\text{r}\}}v^2_{\rho}\kel{g}_{\text{b},\rho}(\omega_n;\epsilon)\delta_{mn}
\end{equation}
with $v_{\text{l}/\text{r}}$ as lead-layer hybridization, $\omega_n \equiv \omega + n\Omega$, $n\in\mathbbm{Z}$, and 
\begin{equation}\label{eq:inv_non-int_lat_GF_comps}
\begin{split}
\left[ g^{-1}_{0}(\omega_n;\epsilon,\overline{\epsilon})) \right]^{\text{R}}_{mn} & = \left( \omega_n + i0^{+} -\varepsilon_{\text{c}} \right)\delta_{mn} - \varepsilon_{mn}(\epsilon,\overline{\epsilon}), \\
\left[ g^{-1}_{0}(\omega_n;\epsilon,\overline{\epsilon})) \right]^{\text{K}}_{mn} & = 0
\end{split}
\end{equation}
as the noninteracting Green's function of the isolated layer, whose inverse Keldysh component is negligible in the steady state. The Floquet dispersion relation $\varepsilon_{mn}$ for the periodic field in a hyper-cubic lattice is~\cite{ts.ok.08} 
\begin{equation}
\label{eq:Floquet_disp}
\varepsilon_{mn}(\epsilon,\overline{\epsilon}) = \begin{cases}
\epsilon J_{m-n}(\aamp) & m-n:\text{even}, \\
\ii\overline{\epsilon}J_{m-n}(\aamp) & m-n:\text{odd},
\end{cases}
\end{equation} 
where $J_n$ denotes the $n$-th order Bessel function of the first kind, with the argument $\aamp$ defined in Sec.~\ref{sec:Model}. The  GF $\kel{g}_{\text{b},\text{l}/\text{r}}$ on the $d$-dimensional boundary of the decoupled leads,  consisting of a single layer coupled to a semi-infinite reservoir in the wide-band limit with broadening $\Gamma_{\text{l}/\text{r}}$ (cf. Fig.~\ref{fig:setup}) is given by
\begin{align}
g^{\text{R}}_{\text{b},\text{l}/\text{r}}(\omega;\epsilon) &= \frac{1}{\omega-\varepsilon_{\text{l}/\text{r}}(\epsilon)+\ii\Gamma_{\text{l}/\text{r}}},\label{eq:retarded_bath_GF} \\
g^{\text{K}}_{\text{b},\text{l}/\text{r}}(\omega;\epsilon) &= 2\ii \iim[g_{\text{b},\text{l}/\text{r}}^{\text{R}}(\omega;\epsilon)][1-2f(\omega,\mu_{\text{l}/\text{r}},\beta)], \label{eq:keldysh_bath_GF}
\end{align} 
where $\varepsilon_{\text{l}/\text{r}}(\epsilon)=\varepsilon_{\text{l}/\text{r}} +\frac{t_{\text{l}/\text{r}}}{t^\ast}\epsilon$ is the dispersion, $f(\omega,\mu_{\text{l}/\text{r}},\beta)=1/[\ee^{\beta(\omega-\mu_{\text{l}/\text{r}})}+1]$ the Fermi-Dirac distribution function at inverse temperature $\beta \equiv 1/T$, $\varepsilon_{\text{l}/\text{r}}$ the onsite energy and $t_{\text{l}/\text{r}}$ the hopping within the $d$-dimensional layer of the lead.  
  
The electron SE $\kel{\mat{\Sigma}}$ is obtained from F-DMFT and therefore, in this approximation, independent of the crystal momentum, i.e. $\kel{\mat{\Sigma}}(\omega;\epsilon,\overline{\epsilon}) \simeq \kel{\mat{\Sigma}}(\omega)$. Further details are given in Sec.~\ref{sec:FDMFT_implementation}. In the same spirit, the e-ph SE is included locally as $\kel{\mat{\Sigma}}_{\text{e-ph}}(\omega;\epsilon,\overline{\epsilon}) \simeq \kel{\mat{\Sigma}}_{\text{e-ph}}(\omega)$. In terms of the Keldysh contour time arguments $z$ and $z^\prime$, it has the form      
\begin{equation}\label{eq:backbone_e-ph_SE}
\Sigma_{\text{e-ph}}(z,z^{\prime}) = \ii g^{2} G(z,z^{\prime}) D_{\text{ph}}(z,z^{\prime}).
\end{equation}
The Keldysh components of the non-interacting phonon GF $\kel{D}_{\text{ph}}(t,t^{\prime})$ are given by~\cite{ao.ts.14}
\begin{align}\label{eq:Ph_Prop_time}
\begin{split} 
D^{\text{R}}_{\text{ph}}(t,t^{\prime}) & = -\ii \theta(t-t^{\prime}) \int \dd\omega \ \ee^{-\ii\omega\left(t-t^{\prime}\right)} A_{\text{ph}}(\omega), \\
D^{>}_{\text{ph}}(t,t^{\prime}) & = -\ii \int \dd\omega \ \ee^{-\ii\omega\left(t-t^{\prime}\right)} A_{\text{ph}}(\omega) \left[ 1 + b(\omega) \right] \\
D^{<}_{\text{ph}}(t,t^{\prime}) & = -\ii \int \dd\omega \ \ee^{-\ii\omega\left(t-t^{\prime}\right)} A_{\text{ph}}(\omega) \ b(\omega), \\
\end{split}, 
\end{align}
where $b(\omega)=1/(\ee^{\beta\omega}-1)$ is the Bose-Einstein distribution function at inverse temperature $\beta$. We consider acoustic phonons, with spectral function $A_{\text{ph}}(\omega) = (\omega/\omega^{2}_{\text{ph}}) \ee^{-| \omega|/\omega_{\text{ph}}}$,  $\omega_{\text{ph}}$ being a soft cutoff frequency~\cite{pi.li.21}. The retarded and Keldysh components of the e-ph SE are easily extracted from  Eq.~\eqref{eq:backbone_e-ph_SE} and can be found in Ref.~\cite{ma.ga.22}. 
     
\subsection{Floquet DMFT}
\label{sec:FDMFT_implementation}

We compute the electron SE in the Dyson equation~\eqref{eq:FullDysonEq} using DMFT~\cite{me.vo.89,ge.ko.92,ge.ko.96}, and in particular its nonequilibrium Floquet extension F-DMFT~\cite{ts.ok.08,sc.mo.02u,jo.fr.08}. In DMFT the crystal momentum dependence of the electron SE is neglected, i.e. $\kel{\mat{\Sigma}}(\omega,\epsilon,\overline{\epsilon}) \to \kel{\mat{\Sigma}}(\omega)$. This allows us to map the original lattice problem onto a single-site impurity model with a bath hybridization function $\kel{\mat\Delta}(\omega)$ encoding the effect of all other lattice sites. 

For completeness, we now briefly describe the self-consistency F-DMFT scheme used. (i) We start from an initial guess for the electron SE $\kel{\mat{\Sigma}}(\omega)$ and set $\kel{\mat \Sigma}_{\text{e-ph}}(\omega)=0$. (ii) Then we compute the local electron GF as 
\begin{equation}\label{eq:Lat_LocGF}
\begin{split}
\kel{\mat G}_{\text{loc}}(\omega) &= \int \dd\epsilon \int \dd\overline{\epsilon} \ \rho(\epsilon,\overline{\epsilon}) \times \\ 
&\times \left[\kel{\mat G}^{-1}_{0}(\omega,\epsilon,\overline{\epsilon}) - \kel{\mat\Sigma}(\omega) - \kel{\mat \Sigma}_{\text{e-ph}}(\omega) \right]^{-1}.
\end{split}
\end{equation}
(iii) Using Eq.~\eqref{eq:backbone_e-ph_SE}, we obtain the phonon contribution to the e-ph SE $\kel{\mat{\Sigma}}_{\text{e-ph}}(\omega)$. (iv) The problem is mapped onto a single impurity plus bath, whose hybridization function is given by
\begin{equation}\label{eq:imp_Dyson_eq}
\kel{\mat{\Delta}}(\omega) = \kel{\mat{g}}^{-1}_{0,\text{site}}(\omega) - \kel{\mat{G}}^{-1}_{\text{loc}}(\omega) - \kel{\mat{\Sigma}}(\omega),
\end{equation}   
where $\kel{\mat{g}}^{-1}_{0,\text{site}}(\omega)$ is defined as in Eq.~\eqref{eq:inv_non-int_lat_GF_comps} with $\varepsilon_{mn}(\epsilon,\overline{\epsilon})=0$. (v) The nonequilibrium many-body impurity problem is solved according to the procedure described below, leading to the new $\kel{\mat{\Sigma}}(\omega)$. (vi) We insert the electron and e-ph SEs into step (ii) and iterate the steps (ii)-(vi) until convergence.

Eq.~\eqref{eq:imp_Dyson_eq} gives a bath hybridization function $\kel{\mat\Delta}(\omega)$ whose periodic time-dependence is encoded in its Floquet structure. This means that, in principle, one should solve a time-periodic impurity problem leading to a non-diagonal SE. However, in Ref.~\cite{so.do.18} it was argued and shown that for the considered electric field amplitudes, off-diagonal terms in the electron SE $\kel{\mat{\Sigma}}(\omega)$ can be safely neglected with respect to the diagonal ones. We follow this argument and adopt here this Floquet-diagonal self-energy approximation (FDSA) as well. In addition, it turns out that also for the e-ph SE $\kel{\mat{\Sigma}}_{\text{e-ph}}(\omega)$ this approximation is justified~\footnote{More specifically, we verified that in the parameter range in which FDSA is justified for the electron SE $\kel{\mat{\Sigma}}(\omega)$ (Sec.~\ref{sec:results}), the off-diagonal terms of the e-ph SE $\kel{\mat{\Sigma}}_{\text{e-ph}}(\omega)$ are sufficiently suppressed}. Consequently, we solve a nonequilibrium stationary impurity problem, by considering only the $(0,0)$-Floquet matrix element of all the quantities in Eq.~\eqref{eq:imp_Dyson_eq}. The other diagonal components of the SEs are then reconstructed by using the property $\kel{\Sigma}_{mm}(\omega) = \kel{\Sigma}_{00}(\omega+m\Omega)$. 

In order to solve the many-body problem in step (v) of the DMFT self-consistent loop, we use the auxiliary master equation approach (AMEA)~\cite{ar.kn.13,do.nu.14,do.ga.15,do.so.17}, which we now briefly summarize. Therein, we map the impurity problem onto an auxiliary open quantum system (AOQS) consisting of a finite number of bath sites $N_{\text{B}}$ attached to Markovian reservoirs described by the Lindblad equation. The  hybridization function $\kel{\Delta}_{\text{aux}}(\omega)$ of this AOQS is obtained by fitting the original DMFT one. The key point is that the many-body problem of this AOQS can be solved exactly using standard many-body-diagonalization methods, as long as  $N_{\text{B}}$ is small. The accuracy of this solution is set by the difference between $\kel{\Delta}_{\text{aux}}$ and $\kel{\Delta}$, which  becomes exponentially small with increasing $N_{\text{B}}$.

\subsection{Physical quantities}
\label{sec:observables}  

To study direct excitation and II effects in this system, we consider the following time-averaged physical quantities.

The photocurrent flowing from the left fermion lead to the right one, passing through the correlated layer is given by two equivalent expressions, adapted from Ref.~\cite{so.do.18}
\begin{align}
j_{\text{l}\rightarrow\text{r}}&=v^2\int_{-\Omega/2}^{\Omega/2}\frac{\dd\omega}{2\pi} \int \dd\epsilon \int \dd\overline{\epsilon} \rho(\epsilon,\overline{\epsilon}) \rre\Tr(\mat{J})\label{eq: Current furmula_Omega}\\
&= v^2\int_{-\infty}^{+\infty}\frac{\dd\omega}{2\pi} \int \dd\epsilon \int \dd\overline{\epsilon} \rho(\epsilon,\overline{\epsilon}) \rre(J_{00}),\label{eq: Current furmula_00}
\end{align}
where
\begin{equation}
  \mat{J}=\left[\mat{G}^{\text{R}}(\mat{g}_{\text{b},\text{l}}^{\text{K}}-\mat{g}_{\text{b},\text{r}}^{\text{K}})+\mat{G}^{\text{K}}(\mat{g}_{\text{b},\text{l}}^{\text{A}}-\mat{g}_{\text{b},\text{r}}^{\text{A}})\right].
\end{equation}
In our case, it is convenient to use Eq.~\eqref{eq: Current furmula_00}. 

The local electronic spectral function (DOS) reads
\begin{equation}
\label{eq:el_spectral_function}
 A(\omega)=-\frac{1}{\pi}\iim[G_{\text{loc},00}^{\text{R}}(\omega)],
\end{equation}
where $G^{\text{R}}_{\text{loc},00}$ is the time-averaged retarded component of the GF given in Eq.~\eqref{eq:Lat_LocGF}. Combining it with the time-averaged Keldysh component gives the occupation function
\begin{equation}
N(\omega) = \frac{1}{4\pi} [\iim(G^{\text{K}}_{\text{loc},00}(\omega))-2\iim(G^{\text{R}}_{\text{loc},00}(\omega))].
\end{equation}

\section{Results}
\label{sec:results}          
     
\begin{figure}[t]    
\includegraphics[width=0.8\linewidth]{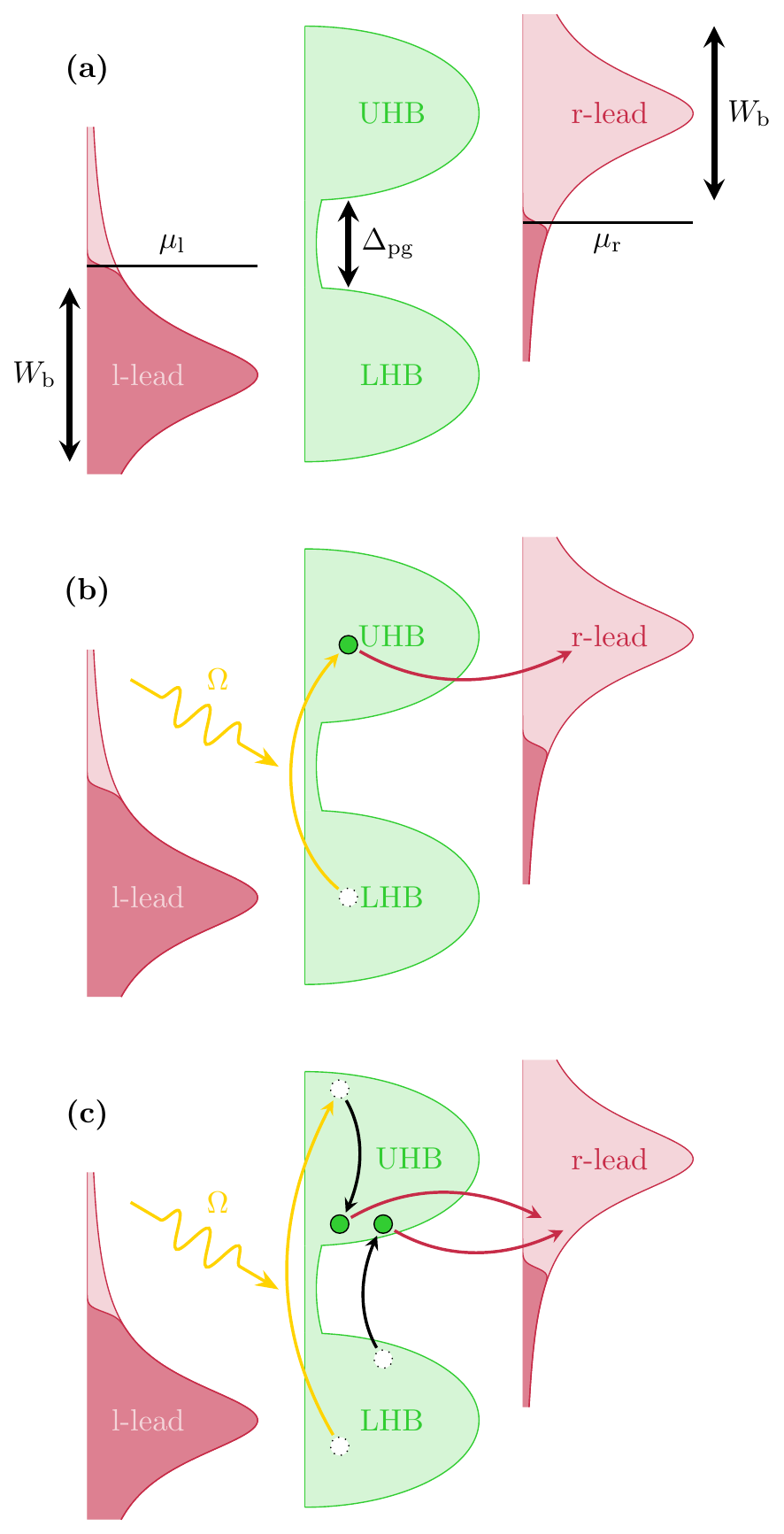}
\caption{(Color online) Schematic representation of the processes occuring in the model considered. Red bands on the left and right describe the leads' DOS, where the dark color highlights occupied states and the light color empty states. Green bands are the LHB and UHB of the central layer. The vertical axis represents the energy. Due to the hybridization with the metallic leads only a pseudogap $\Delta_{\text{pg}}$ is present between the Hubbard bands. Panel (a) sketches the quantities introduced at the beginning of Sec.~\ref{sec:results}. Panel (b) illustrates a direct excitation process, in which an electron is excited by a photon with energy $\Omega$ (yellow arrow) to the UHB and escapes into the right lead (red arrow). Panel (c) displays an II process, where the photoexcited electron in the UHB excites a second electron from LHB to UHB (black arrows) and both escape into the right lead. In the nonequilibrium steady state considered here, only processes recovering the initial configuration are allowed.} 
\label{fig:energy_setup}
\end{figure}

In order to study II in a Mott insulating layer, we adjust the bands to achieve the energy scheme shown in Fig.~\ref{fig:energy_setup}(a). We follow Ref.~\cite{so.do.18} and take $U=12$ and unless stated otherwise $E_0=2$. The fermion leads and the acoustic phonons have temperature $T=0.02$~\footnote{The value chosen for the temperature is much smaller than the other characteristic energy scales of the system, e.g. $\omega_{\text{ph}}$, $U$ and $W_{\text{b}}$.}. We consider a particle-hole symmetric system with $\Gamma_{\text{l}}=\Gamma_{\text{r}}=0.37$, $t_{\text{l}}=t_{\text{r}}=1.7$, $\varepsilon_{\text{l}/\text{r}}=\mp 6$ and $v_{\text{l}}=v_{\text{r}}=v$. The parameters $t_{\text{l/r}}$, $\varepsilon_{\text{l/r}}$, and  $\Gamma_{\text{l/r}}$ are chosen such that the leads' DOS approximately overlaps with the Hubbard bands, i.e. they exhibit a \textquoteleft bandwidth\textquoteright~\footnote{Notice that the support of the leads' spectra are formally infinite. $W_{\text{b}}$ is the value at which the spectrum gets  suppressed.} $W_{\text{b}}\approx 8$ and are centered at the same position. The chemical potentials are set to $\mu_{\text{l}/\text{r}} =\mp 1$, so that a current from left to right is produced by taking energy from the driving. Due to the hybridization with the leads, the local DOS of the Hubbard layer only features a pseudogap
 $\Delta_{\text{pg}} \approx 4$. All simulations are carried out with the dimensionless factor $\alpha\equiv t_{\text{resc}}E_0/\Omega^{2}<0.5$ for which the FDSA is justified~\footnote{In Ref.~\cite{so.do.18} the factor is defined as $\alpha=t_{c}E_0/\Omega^{2}$ and $t_{\text{c}}=1$. Setting $t_{\text{resc}}=1$, we define for consistency $\alpha$ as in the main text.}. Whenever the electron-phonon interaction is included, we take $g=0.8$ and unless stated otherwise $\omega_{\text{ph}}=0.1$.  

After some preliminary considerations in Sec.~\ref{sec:explicative_scheme}, we discuss the occurrence of II at different $v$ first without (Sec.~\ref{sec:E0_2_only_electrons}) and then with coupling to acoustic phonons (Sec.~\ref{sec:E0_2_electrons_phonons}). 

\subsection{Qualitative energy considerations and physical processes}
\label{sec:explicative_scheme}
 
To infer the conditions necessary for II, we consider the  scheme in Fig.~\ref{fig:energy_setup}. Thereby, we partially follow Ref.~\cite{so.do.18} for the electron-only (EO) case and extend the analysis to the case with e-ph interaction. In comparison with Ref.~\cite{so.do.18}, here the wide-band leads describe more realistically a metallic structure. This makes the identification of II processes more subtle, as discussed below. As in Ref.~\cite{so.do.18} we bypass transient behavior and consider directly the Floquet steady state. In our analysis, we neglect higher-order processes such as multiparticle scattering processes. 

In order to observe II, the bandwidth of the UHB has to be at least twice the pseudogap~\footnote{Due to the hybridization with the metallic leads only a pseudogap $\Delta_{\text{pg}}$ is present between the LHB and the UHB.} size $\Delta_{\text{pg}}$. Only in this case, the photoexcited electron can acquire enough energy to excite a second one across the pseudogap.

\subsubsection{EO system}
\label{sec:scheme_only_electrons}  
 
\begin{itemize}
\item For $\Omega<\Delta_{\text{pg}}$, 
we expect a strong suppression of the current~\footnote{The remaining leak current originates either from multiple-photon absorption processes which are strongly suppressed, or from the absence of a true gap.} and we do not expect any current if the DOS of the correlated layer has a true gap.
\item For $\Delta_{\text{pg}}<\Omega<\Delta_{\text{pg}}+2W_{\text{b}}$ as shown in Fig.~\ref{fig:energy_setup}(b), an electron coming from the left lead into the LHB is photoexcited to the UHB and can escape directly into the right lead without additional scattering.  Such processes are often referred to as direct excitations (DEs).
\item For $2\Delta_{\text{pg}}<\Omega<\Delta_{\text{pg}}+2W_{\text{b}}$ as shown in Fig.~\ref{fig:energy_setup}(c), a photoexcited electron in the UHB can excite via II a second electron from LHB to UHB, before both escape into the right lead.
\item For $\Omega>\Delta_{\text{pg}}+2W_{\text{b}}$, there are no final states available for a photoexcited electron and we do not expect the transition to occur. 
\end{itemize}
Notice that in the energy window $\Delta_{\text{pg}}<\Omega<2\Delta_{\text{pg}}$ the only scattering processes taking place are DEs, while for $2\Delta_{\text{pg}}<\Omega<\Delta_{\text{pg}}+2W_{\text{b}}$, both DE and II can occur. This is in contrast to the case of Ref.~\cite{so.do.18}, in which the ad-hoc narrow leads' bands only allow for II processes in a certain $\Omega$-range. For this reason, and due to the fact that the boundaries of these energy ranges are not strict, it is more difficult to disentangle these physical processes in the present case.

\subsubsection{Inclusion of e-ph interactions}
\label{sec:scheme_electrons_phonons}  

Upon inclusion of e-ph scattering, the outlined scheme remains valid except for the fact that  the pseudogap $\Delta_{\text{pg}}$ is slightly modified. As shown in Sec.~\ref{sec:E0_2_electrons_phonons}, the phonons broaden the DOS of the correlated layer, which slightly fills and shrinks the pseudogap.
 
\begin{figure*}[t]
\includegraphics[width=0.7\textwidth]{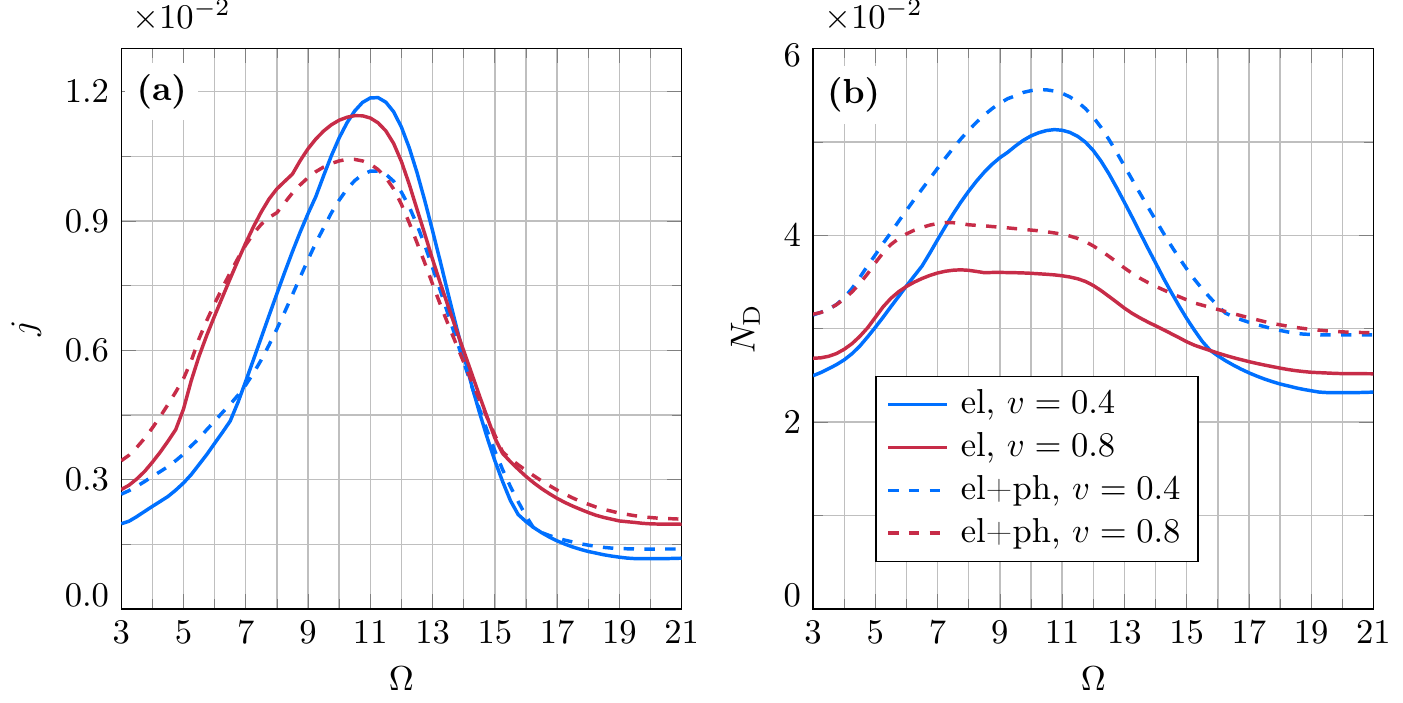}
\caption{(Color online) 
 (a) Time-averaged steady state current $j$ and (b) double occupancy $N_{\text{D}}$ plotted as a function of the driving frequency $\Omega$. Results are shown for different values of the lead-layer hybridization $v$ as well as without and with e-ph interaction.
 Default parameters are specified at the beginning of Sec.~\ref{sec:results}.} 
\label{fig:j_vs_omega_mu1_v_0.4_0.8_eph}
\end{figure*}

\subsection{EO system}
\label{sec:E0_2_only_electrons}   
 
We start by analyzing the case without e-ph interaction and address the behavior of the physical quantities as functions of $\Omega$ for different values of the lead-layer hybridization.

In Fig.~\ref{fig:j_vs_omega_mu1_v_0.4_0.8_eph}(a), the photocurrent $j$ increases as a function of the driving frequency $\Omega$ within the range $4\lesssim\Omega\lesssim8$. According to the discussion in Sec.~\ref{sec:scheme_only_electrons} (cf.  Fig.~\ref{fig:energy_setup}), this $\Omega$-range is expected to allow only for DEs. The current in this range  increases with increasing $v$. Similarly, the double occupation shown in Fig.~\ref{fig:j_vs_omega_mu1_v_0.4_0.8_eph}(b) increases as a function of the driving frequency~\footnote{Notice that the increase of $N_{\text{D}}$ in Fig.~\ref{fig:j_vs_omega_mu1_v_0.4_0.8_eph}(b) as compared to Ref.~\cite{so.do.18} is not as steep and large in magnitude. This is due to our setup, where every photoexcited carrier can escape into the right lead.}.

\begin{figure}[b] 
\includegraphics[width=\linewidth]{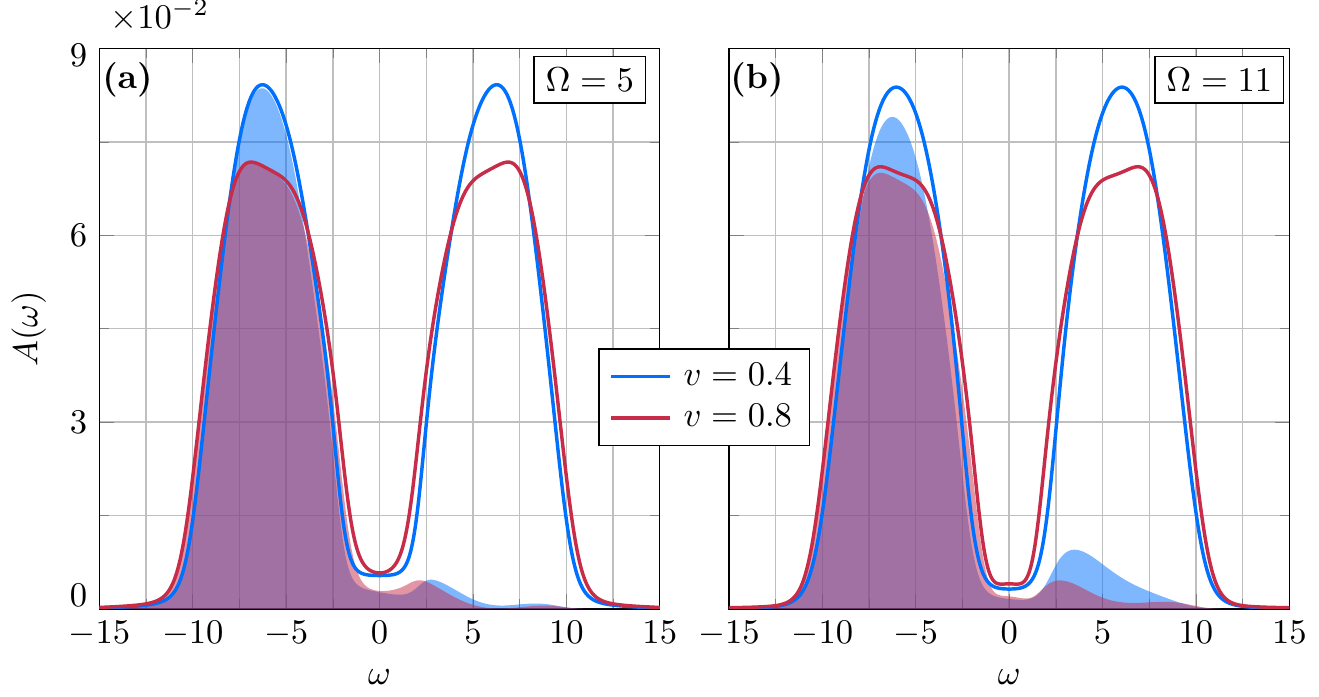}
\caption{(Color online) Spectral function $A(\omega)$ (solid line) and occupation function $N(\omega)$ (shaded area) at (a) $\Omega=5$ and (b) $\Omega=11$ for $v=0.4$ and $0.8$ for the EO system. Default parameters are specified at the beginning of Sec.~\ref{sec:results}.}
\label{fig:spec_filling_mu1_v_0.4_0.8_O_5_11_e}
\end{figure} 

\begin{figure}[b]  
\includegraphics[width=\linewidth]{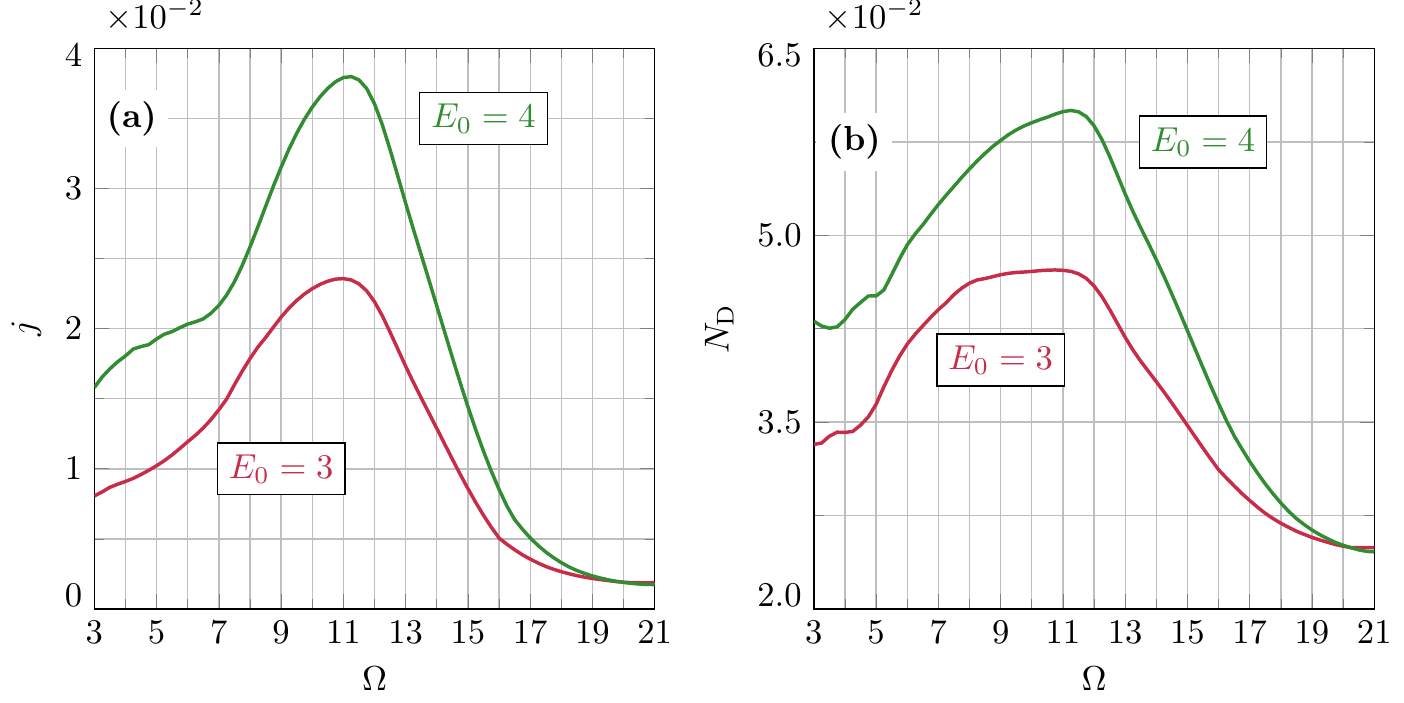} 
\caption{(Color online) (a) Time-averaged steady state current $j$ and (b) double occupancy $N_{\text{D}}$ as a function of $\Omega$, at $v=0.8$ for $E_0=3,4$ for the EO system. Default parameters are specified at the beginning of Sec.~\ref{sec:results}.} 
\label{fig:j_vs_omega_mu1_v_0.8_e_E0_3_4}
\end{figure}

For $\Omega\gtrsim8$, the current increases further until it reaches its peak at $\Omega\approx 11$. In this range both DE and II can occur in principle. The height of the peak is approximately the same for both $v$, despite the hybridizations differ by a factor of two. The double occupation in this range shows a similar behavior as the current for $v=0.4$ with a peak around $\Omega\approx 11$, while it exhibits a plateau for $v=0.8$. For $v=0.4$ the current curve is characterized by a substantial increase in slope around $\Omega \approx 7$, which cannot be observed for $v=0.8$. This, together with the comparable magnitude of the current maxima at $\Omega\approx 11$ and the behavior of the double occupation strongly suggests that for $v=0.4$ a substantial amount of II processes take place, while these are absent or negligible for $v=0.8$. Here, the plateau in $N_{\text{D}}$ for $v=0.8$ suggests that photoexcited electrons arriving in the UHB quickly escape to the right fermionic lead and do not have time to induce II processes.

The spectral and occupation functions shown in Fig.~\ref{fig:spec_filling_mu1_v_0.4_0.8_O_5_11_e} corroborate this hypothesis. For $\Omega=5$, in the DE range, the occupation function in the UHB is almost the same for both $v$, while for $\Omega=11$ where the peak in the current occurs, the occupation of the UHB for $v=0.4$ is substantially larger, in line with the behavior of $N_{\text{D}}$. Further increasing $\Omega$ produces excitations near the border of the UHB where the DOS is reduced and thus  current and double occupation decrease~\footnote{The current $j$ does not approach zero as expected for $\Omega>20$ because of the background current as discussed in Ref.~\cite{so.do.18}.}.

This behavior is affected by the electric field amplitude $E_0$ as illustrated in Fig.~\ref{fig:j_vs_omega_mu1_v_0.8_e_E0_3_4}(a). For $E_0=4$ the current develops a sudden increase in slope around $\Omega\approx7$, as well as a pronounced peak around $\Omega\approx 11$ for the larger hybridization $v=0.8$. Both features start developing already at $E_0=3$. This tendency, corroborated by the behavior of the double occupation displayed in Fig.~\ref{fig:j_vs_omega_mu1_v_0.8_e_E0_3_4}(b), suggests an  important role of II also for larger values of $v$, provided the field amplitude is strong enough.

\subsection{Inclusion of e-ph interactions}
\label{sec:E0_2_electrons_phonons} 

We now discuss the effect of coupling to acoustic phonons on the results presented thus far. 
  
The current $j$ in presence of phonons shown in Fig.~\ref{fig:j_vs_omega_mu1_v_0.4_0.8_eph}(a) is slighty larger than the EO one for $\Omega\lesssim7$. This effect is reduced with increasing $\Omega$ until a crossing occurs at $\Omega\approx7$ where the EO current overtakes. Between $7\lesssim\Omega\lesssim14$, the current in the presence of phonons exhibits the same qualitative behavior as the EO one, reaching its maximum at $\Omega\approx 11$. However the current magnitude is suppressed, especially around the peak. For $\Omega\gtrsim14$ the current in presence of phonons drops as for the EO case.

\begin{figure}[b]
\includegraphics[width=\linewidth]{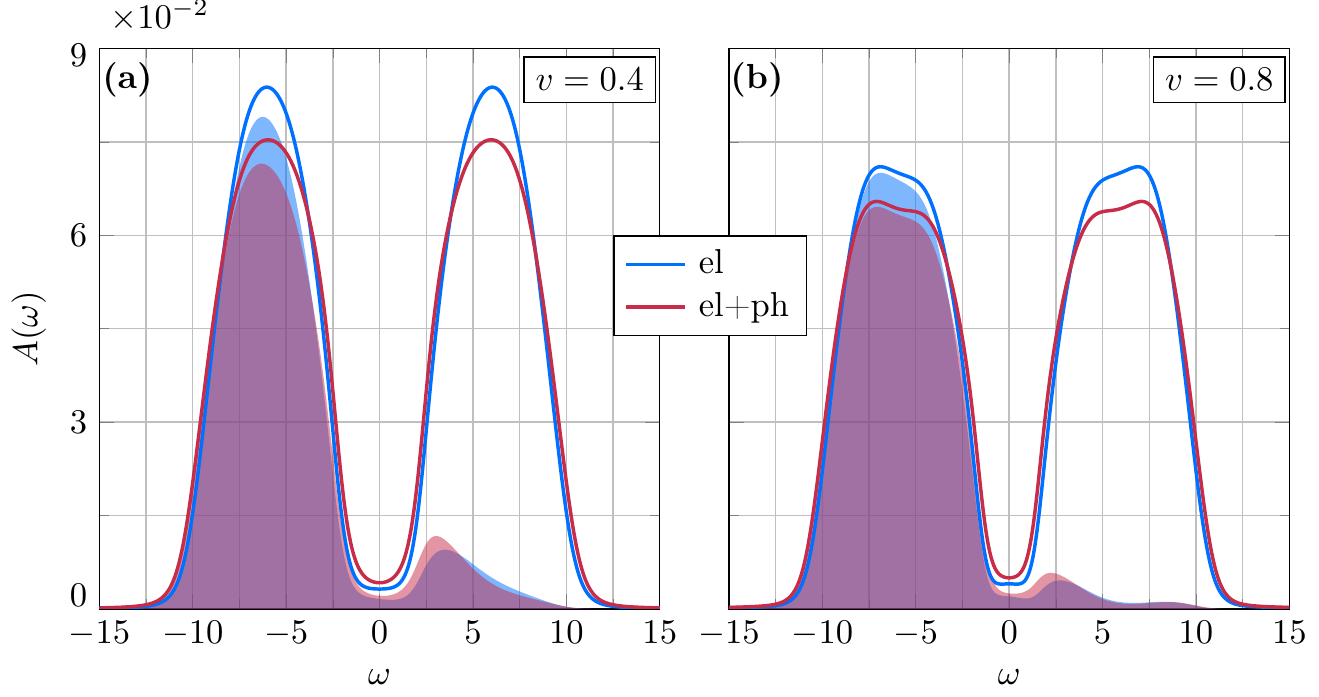}
\caption{(Color online) Spectral function $A(\omega)$ (solid line) and occupation function $N(\omega)$ (shaded area) at $\Omega=11$ without and with e-ph interaction, for (a) $v=0.4$ and (b) $v=0.8$. Default parameters are specified at the beginning of Sec.~\ref{sec:results}.} 
\label{fig:spec_filling_mu1_v_0.4_0.8_O_11_eph}
\end{figure}

The double occupation $N_{\text{D}}$  shown in Fig.~\ref{fig:j_vs_omega_mu1_v_0.4_0.8_eph}(b) is larger than and follows the behavior of its EO counterpart within the entire $\Omega$-range considered. The spectral and occupation functions  shown in Fig.~\ref{fig:spec_filling_mu1_v_0.4_0.8_O_11_eph} display a redistribution of spectral weight from the peaks towards the edges of the bands, thus reducing the pseudogap. Consequently, the occupation of positive energy states is slightly shifted to the bottom of the UHB compared to the EO case, as evidenced by the occupation function. The spectral function in presence of phonons is broadened and tends to fill the pseudogap $\Delta_{\text{pg}}$ as anticipated in Sec.~\ref{sec:scheme_electrons_phonons}. This reduction of the pseudogap explains why the current in Fig.~\ref{fig:j_vs_omega_mu1_v_0.4_0.8_eph}(a) is slightly larger in presence of phonons for $\Omega\lesssim7$. With a smaller pseudogap and more states at its edges, more electrons can be directly excited with a given driving frequency $\Omega$ thereby increasing the photocurrent. On the other hand, the suppression of the current by e-ph interaction in the range $7\lesssim\Omega\lesssim14$ is due to the dissipation-induced reduction of spectral weight at frequencies $|\omega| \gtrsim 3.5$, see Fig. \ref{fig:spec_filling_mu1_v_0.4_0.8_O_11_eph}. 

The increase of the double occupation $N_{\text{D}}$ seen in Fig.~\ref{fig:j_vs_omega_mu1_v_0.4_0.8_eph}(b) for the case with phonons can already be inferred from the occupation function $N(\omega)$ at positive frequencies depicted in Fig.~\ref{fig:spec_filling_mu1_v_0.4_0.8_O_11_eph}.

\begin{figure}[t] 
\includegraphics[width=\linewidth]{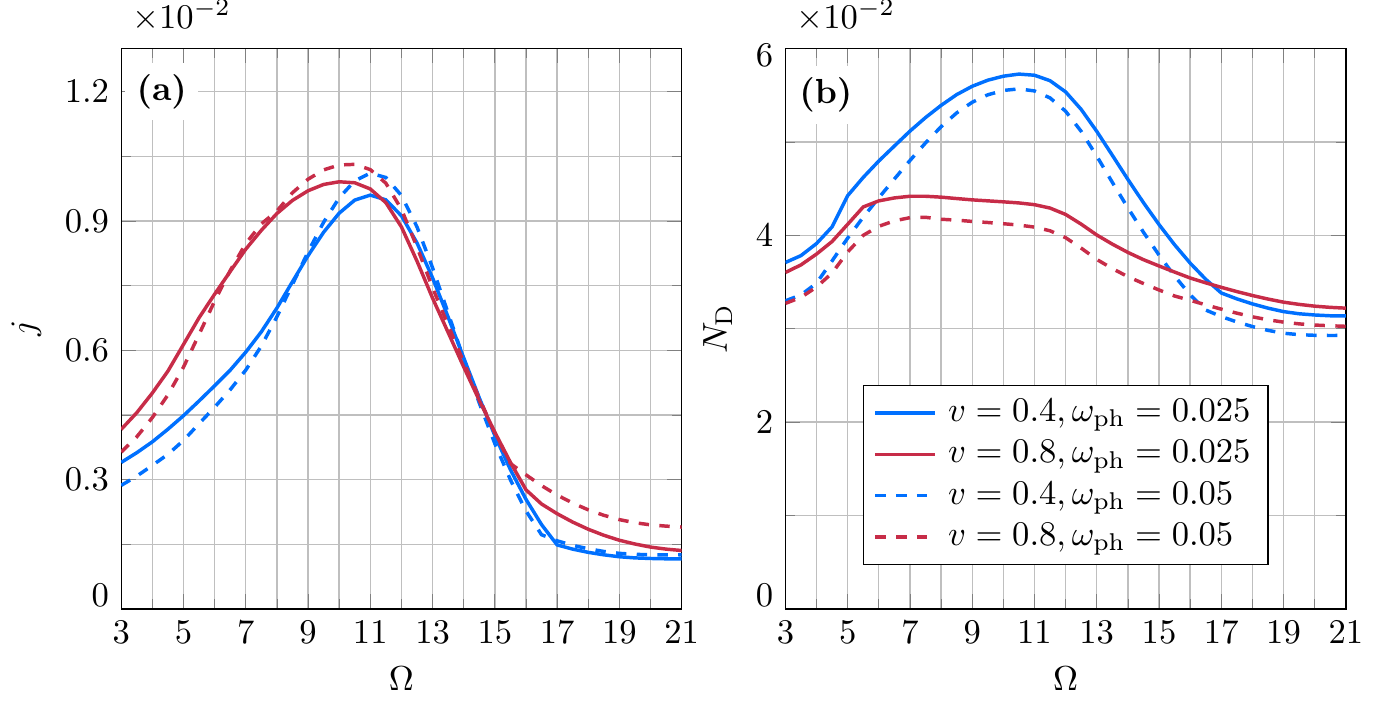}
\caption{(Color online) (a) Time-averaged steady state current $j$ and (b) double occupancy $N_{\text{D}}$ as a function of $\Omega$, for $v=0.4, 0.8$, at phonon cutoff frequencies $\omega_{\text{ph}}=0.025, 0.05$. Default parameters are specified at the beginning of Sec.~\ref{sec:results}.} 
\label{fig:j_vs_omega_mu1_v_0.4_0.8_E0_2_sweep_omegaph}
\end{figure}

Summarizing, phonons slightly enhance the current for almost all driving frequencies $\Omega\lesssim 7$ for which only DEs take place. They suppress the current in the high-frequency range $7\lesssim\Omega\lesssim14$.

This behavior is qualitatively valid for different values of the soft cutoff phonon frequency $\omega_{\text{ph}}$. It can be seen in Fig.~\ref{fig:j_vs_omega_mu1_v_0.4_0.8_E0_2_sweep_omegaph}, which shows that the impact of acoustic phonons on current and double occupation is slightly boosted when decreasing $\omega_{\text{ph}}$~\footnote{Reducing $\omega_{\text{ph}}$ has two effects. First, the phonon spectral function $A_{\text{ph}}(\omega)$ exhibits more weight at low frequencies around $\omega\approx\omega_{\text{ph}}$. Second, it restricts the maximum value of the reciprocal lattice vector $\vec{q}_{\text{max}}$ via $\omega_{\text{ph}}=\omega(\vec{q}_{\text{max}})$~\cite{ma.ga.22}, meaning that short-wavelength phonons are suppressed.}. 
     
\begin{figure}[b] 
\includegraphics[width=\linewidth]{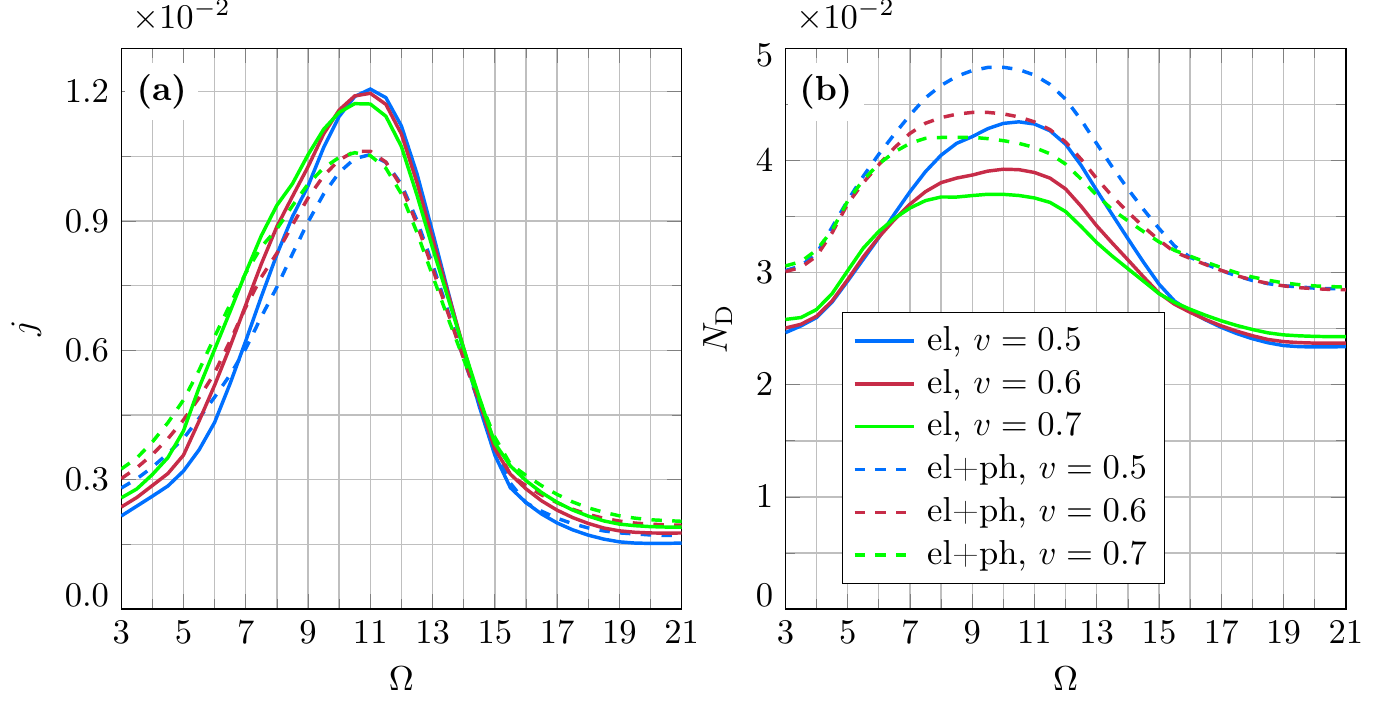}
\caption{(Color online) (a) Time-averaged steady state current $j$ and (b) double occupancy $N_{\text{D}}$ as a function of $\Omega$, for $v=0.5,0.6,0.7$, without and with e-ph interaction. Default parameters are specified at the beginning of Sec.~\ref{sec:results}.} 
\label{fig:j_vs_omega_mu1_sweep_v_E0_2_eph}
\end{figure} 
     
The results above suggest that the influence of acoustic phonons on the electronic scattering processes does not depend significantly on $v$. In other words, the effect of phonons does not depend significantly on the rate at which carriers are injected into and removed from the layer. Fig.~\ref{fig:j_vs_omega_mu1_sweep_v_E0_2_eph} confirms this behavior for intermediate lead-layer hybridizations $v=0.5,0.6,0.7$. Also the location of the crossing in the current  between the EO and e-ph case is essentially independent of $v$, as can be seen in Fig.~\ref{fig:j_vs_omega_mu1_sweep_v_E0_2_eph}(a).

\section{Conclusion}   
\label{sec:conclusions} 
 
We investigated the influence of fermion leads and phonon dissipation on electron transport and spectral properties of a Mott insulating layer driven to the nonequilibrium steady state by an external periodic electric field. In order to realize a Mott-based photovoltaic setup between metallic leads, we considered a correlated layer coupled to acoustic phonons, located between wide-band fermion leads. We studied the influence of the strength of the coupling to the leads on the scattering process occurring at different driving frequencies and observed how the dissipation by acoustic phonons influences the photocurrent and the double occupation. We found evidence of a significant amount of II processes leading to a photocurrent peak at small hybridizations, while these are suppressed  for larger hybridizations. Dissipation via acoustic phonons slightly boosts the photocurrent for small driving frequencies and suppresses it at larger ones in the vicinity of the main peak. The effect of phonons is not affected significantly by changes in the lead-layer hybridization.

One should comment on the validity of DMFT, which should be accurate for large dimensions only, for the present quasi two-dimensional (2D) setup. As a matter of fact, the hybridization to the leads introduces a dissipation which suppresses 2D-coherence and thus makes nonlocal correlations beyond DMFT less relevant.
On the other hand, it has been shown that antiferromagnetic (even short-range) correlations may play an important role in the spreading of photoexcited carriers~\cite{ec.we.14}. 
Taking into account the effect of such nonlocal correlations (see also~\cite{ro.ha.18}) might be an interesting extension, although beyond the scope of the present research.

Experimentally, the electric field amplitude considered in this paper is several orders of magnitude larger than that of the sunlight, therefore our results are more relevant for photoexcitations produced by intense laser pulses (see also the discussion in~\cite{mu.we.18}) rather than for true photovoltaic systems. We did not consider smaller electric field amplitudes due to the slow F-DMFT convergence for such a choice of parameters. This is also the case for weaker hybridization strengths. 
 
On the other hand, one may try to qualitatively extrapolate the discussion at the end of Sec.~\ref{sec:E0_2_only_electrons} in the other direction to smaller electric field amplitudes, for which a smaller hybridization would be necessary in order to sustain II. In a more realistic setup modelling oxide heterostructures~\cite{as.bl.13,pe.be.19}, a regime of effectively small hybridization would be achieved by considering several layers in which photoexcited carriers, separated by an electric field gradient, have more time to induce II processes before escaping into the leads. To address the occurrence of II in such a setup, further extensions such as considering multiple orbitals and impurity scattering should be taken into account. This aspect could be interesting for future studies.    
Finally, a more realistic description of the effects of the electron-phonon interaction on electronic band structure should take into account the crystal momentum dependence of $\kel{\mat{\Sigma}}_{\text{e-ph}}$~\cite{gius.17}.

\acknowledgments  

We thank A. Picano and C. Heil for fruitful discussions. This work was supported by the Austrian Science Fund (Grant No. P 33165-N) and by NaWi Graz. The computational results presented have been obtained using the Vienna Scientific Cluster (VSC) and the D-Cluster Graz.

\bibliographystyle{prsty}   
\bibliography{references_database}

\end{document}